\documentclass[conference]{IEEEtran}
\ifCLASSINFOpdf
   \usepackage[pdftex]{graphicx}
   \DeclareGraphicsExtensions{.pdf,.jpeg,.png}
\else
\fi
\usepackage{tabularx,calc}
\usepackage{color}
\usepackage{tikz,amsmath, amssymb,bm,color}
\usepackage{pgfplots}
\usepackage{environ}
\usetikzlibrary{shapes,arrows}
\usetikzlibrary{calc}

\makeatletter
\newsavebox{\measure@tikzpicture}
\NewEnviron{scaletikzpicturetowidth}[1]{%
	\def\tikz@width{#1}%
	\begin{lrbox}{\measure@tikzpicture}%
		\BODY
	\end{lrbox}%
	\pgfmathparse{#1/\wd\measure@tikzpicture}%
	\BODY
}
\makeatother
\usepackage{ifthen}
\newboolean{compileforpublish}
\setboolean{compileforpublish}{true}
\newboolean{isaccepted}
\setboolean{isaccepted}{false}
\ifthenelse{\boolean{isaccepted}}
{%if
	\newcommand\copyrighttext{%
		\footnotesize \parbox[t]{.11\textwidth}{\copyright{} \the\year~IEEE.} \parbox[t]{.89\textwidth}{Personal use of this material is permitted. Permission from IEEE must be obtained for all other uses, in any current or future media, including reprinting/republishing this material for advertising or promotional purposes, creating new collective works, for resale or redistribution to servers or lists, or reuse of any copyrighted component of this work in other works.}}
}
{%else
	\newcommand\copyrighttext{%
		\footnotesize \centering This work has been submitted to the IEEE for possible publication.\\ Copyright may be transferred without notice, after which this version may no longer be accessible.}
}

\newcommand\copyrightnotice{%
	\ifthenelse{\boolean{compileforpublish}}
	{
		\begin{tikzpicture}[remember picture,overlay]
		\node[anchor=south,yshift=10.5pt] at (current page.south) {\parbox{\dimexpr\textwidth-\fboxsep-\fboxrule\relax}{\copyrighttext}};
		\end{tikzpicture}%
	}
}

\renewcommand{\IEEEtitletopspaceextra}{20pt}

\begin{document}
\title{A System's Perspective Towards an Architecture Framework for Safe Automated Vehicles}

\author{\IEEEauthorblockN{Gerrit Bagschik, Marcus Nolte, Susanne Ernst and Markus Maurer}
\IEEEauthorblockA{Institute of Control Engineering\\
Technische Universit\"at Braunschweig\\
Braunschweig, Germany\\
Email: \{bagschik, nolte, ernst, maurer\}@ifr.ing.tu-bs.de}
}
\maketitle%
\copyrightnotice%

\begin{abstract}
With an increasing degree of automation, automated vehicle systems become more complex in terms of functional components as well as interconnected hardware and software components.
Thus, holistic systems engineering becomes a severe challenge.
Emergent properties like \emph{system safety} are not solely arguable in singular viewpoints such as structural representations of software or electrical wiring (e.g. fault tolerant). 
This states the need to get several viewpoints on a system and describe correspondences between these views in order to enable traceability of emergent system properties.
Today, the most abstract view found in architecture frameworks is a logical description of system functions which structures the system in terms of information flow and functional components.
In this article we extend established system viewpoints towards a capability-based assessment of an automated vehicle and conduct an exemplary safety analysis to derive behavioral safety requirements.
These requirements can afterwards be attributed to different viewpoints in an architecture frameworks and thus be integrated into a development process for automated vehicles.
\end{abstract}

\section{Introduction}

While more and more OEMs, tech-companies, and research facilities demonstrate automated driving, ensuring safety of these vehicles is still an unsolved topic.
Developing safe vehicles is a complex task, involving multiple viewpoints on the system under development.
Koopmann and Wagner point out that 'ensuring the safety of fully autonomous vehicles requires a multidisciplinary approach across all the levels of functional hierarchy' \cite[p. 1]{koopman_autonomous_2017}.
%, from hardware fault tolerance, to resilient machine learning, to cooperating with humans driving conventional vehicles, to validating systems for operation in highly unstructured environments, to appropriate regulatory approaches.''

When involving multiple disciplines in systems engineering, there are several concerns on the system which need to be addressed in architectural viewpoints. 
Those viewpoints have to be translated or connected to each other, not only for identifying and developing all required details of a safe system, but also for creating a holistic systems engineering process.
Systems engineering also puts high demands on traceability to argue a sufficient level of safety, since absolute safety is neither possible nor provable.
Behere et al. \cite{menegatti_architecture_2016} also conclude a workshop with many senior system architects from a broad cross-section of the industry that \emph{correctness by construction} will play a significant role in development of autonomous systems.
\emph{Correctness by construction} can be translated to \emph{safety by design} can in terms of safety and can only be provided by multidisciplinary and cross-domain efforts from academic disciplines \cite{menegatti_architecture_2016}.

The need for a holistic engineering process is also demanded by emergent properties of a complex system.
According to Checkland, \emph{emergence} is ``the principle that entities exhibit properties which are meaningful only when attributed to the whole, not to its parts.'' \cite[p. 314]{checkland_systems_1981}.
Related to safety, this means that safety of an automated vehicle can not solely be analyzed or verified by looking at, for example, a hardware architecture.
Therefore, requirements and concepts for safe vehicles must be derived in a top down development process which incorporates different views on a system at all abstraction levels.
Waymo \cite{waymo_road_2017} recently described multiple aspects of safety: \emph{behavioral safety, functional safety, crash safety, operational safety,} and \emph{non-collision safety}.

\emph{Functional safety} for automated vehicles is addressed by standards like the ISO~26262.
The main focus of functional safety is on reaching low hardware failure rates together with high software quality.
The according safety requirements result from a risk assessment which is solely based on potential failures of the E/E-system.
Functionally safe systems ensure that functional (safety) requirements are correctly implemented according to the demands of the automotive safety integrity level (ASIL) rating.

The question whether a driving function under normal operation behaves in a safe way is not scope of the ISO~26262 and part of \emph{behavioral safety}.
Therefore, efforts like \emph{safety of the intended functionality}\footnote{ISO/WD PAS 21448 - under development} (SOTIF) are currently developed complementary to the ISO~26262 standard.
Additionally, the normative concerns on the behavior of automated vehicles are discussed by other institutions like ethics commissions \cite{federal_ministry_of_transport_and_digital_infrastructure_ethics_2017} and government agencies \cite{path_program_peer_2016}.
An example of the difference between behavioral and functional safety is given by a hazard analysis and risk assessment we conducted for an unmanned protective vehicle \cite{stolte_hazard_2017}.
The unmanned protective vehicle is designed to follow a working vehicle with up to 6.5~mph automated and unmanned on the hard shoulder of a motorway.
During the analysis we, investigated possible scenarios with and without possible failures of E/E-systems.
In scenarios without failures, we found some cases where the responsibilities of the construction workers with respect to the system were unclear.
These cases can lead to mode confusion where the unmanned operation of the protective vehicle could have been activated in unwanted or forbidden scenarios.
Therefore, we defined a feedback on the item definition from the hazard analysis and risk assessment to ensure a safe definition of the driving function under normal operation.
If the components which implement the possible mode confusion were only investigated towards functional safety, the system would contain hazards in high quality software running on fail-safe or fail-operational hardware.
This example shows the necessity of a hierarchical or parallel analysis of the (externally visible) behavior and the (internal) system behavior which are connected through requirements during the development process.
Johansson et al. \cite{watzenig_functional_2017} also argue that functional safety for autonomy\footnote{Since \emph{functional safety for autonomy} is further defined as the system level behavior of all items in certain scenarios, we assume this term to be equivalent to \emph{behavioral safety}.} is not covered by the ISO~26262 standard.
They conclude that the semantics of safety requirements ('What does the autopilot do?' \cite[p. 12]{watzenig_functional_2017}) are strongly connected to the architectural allocation of functions and elements in the system ('Where does the autopilot hide?' \cite[p. 12]{watzenig_functional_2017}).

emph{Operational safety} is meant to analyze how humans and other traffic participants interact with the automated vehicle. 
This viewpoint of safety on the system also defines important requirements but is not the scope of our investigation in this article.
Crash and non-collision safety are more ``classical'' viewpoints on the vehicle which are covered and well investigated in today's development processes.

This article shows an approach how safety requirements derived from \emph{behavioral safety} can be modeled in an capability based viewpoint and afterwards be allocated into a functional system architecture which represents the functional structure of the system. 
Based on the functional system architecture and the modeled requirements from \emph{behavioral safety}, we can conduct an analysis on \emph{functional safety}.
This top-down development of requirements and viewpoints enables holistic systems engineering for safe automated vehicles.

The next section reviews current architectural viewpoints and architecture meta models and formulates research needs.
Afterward, we describe our approach based on an example scenario at a pedestrian crossing.
At the end we, conclude our results and show links to existing architecture frameworks.

\section{Related work}

In this section, we first describe related work in architecture frameworks and afterwards discuss contributions to logical architectures covering safety aspects.

\subsection{Architecture Frameworks}

Multiple viewpoints covering concerns towards a system have been discussed extensively in literature.
The basic need for viewpoints was already stated by Dijskstra as ``separation of concerns'' \cite{dijkstra_ewd_1982} for software intense systems.
The main idea is, that a single architecture or viewpoint cannot cover all concerns on a system.

\begin{figure}[!h]
	\centering
	\includegraphics[width=\columnwidth]{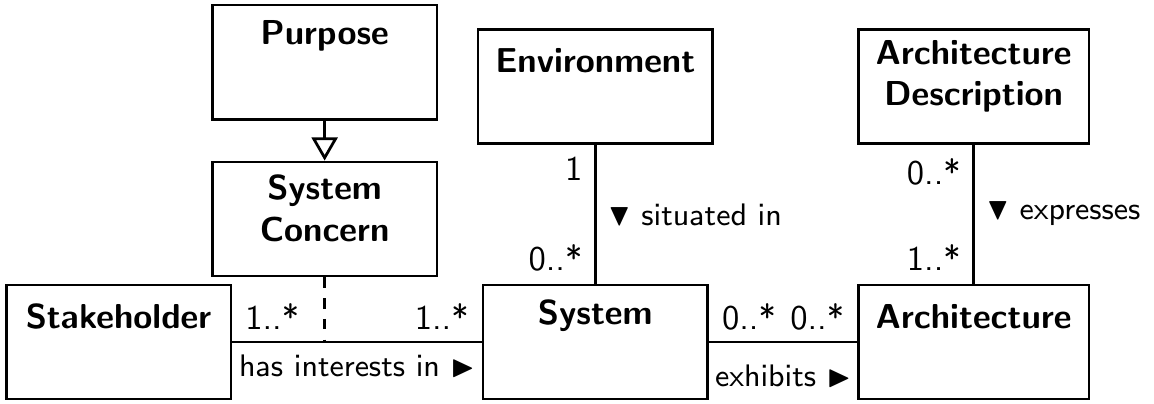}
	\caption{Concepts of the ISO/IEC/IEEE 42010 \cite{ISO_42010_2011}}
	\label{fig:ISO42010}
	
\end{figure}

The ISO/IEC/IEEE 42010 \cite{ISO_42010_2011} standard describes a meta model for architecture frameworks which focuses on a process to identify stakeholders, concerns and viewpoints, and how architecture descriptions shall be formulated.
However, definitions of concrete concerns or viewpoints are not covered by the standard.
The basic concepts and relations between the described elements of the standard are shown in Fig.~\ref{fig:ISO42010}.

A system is working in an environment which can have other systems or be a physical environment.
Automated vehicles will mostly operate in public traffic.
Stakeholders then have concerns on a system.
In this contribution we focus on the safety concern which influences multiple viewpoints as introduced in the previous section.
The system then can be modeled in several architectures which contain requirements resulting from concerns.
An architecture is based on an architecture description which is a blueprint containing elements which can be used.
The standard also describes how different architectures correspond to each other which is important to trace architectural decisions throughout the life cycle of a system. 

Broy et al. \cite{broy_toward_2009} propose a holistic architecture framework for the automotive industry.
The taxonomy is based on a draft version of the later ISO/IEC/IEEE 42010.
The authors propose three levels of abstraction for architecture starting from the most abstract functional architectures.
Functional architectures are detailed by logical architectures and then implemented by technical architectures.
The mandatory viewpoints for the framework are \emph{functional, technical, information, driver/vehicle operations} and \emph{value net}.
Safety is an example of an optional viewpoint which cross cuts the proposed framework.
One of the main motivations to utilize model based development through the architectures is that requirements are neither complete nor static at the beginning of development and have to be analyzed and defined recursively throughout the process.
%This is an important aspect as it holds requirements for the correspondences between architectures. 
Since the framework does not explicitly model or analyze safety of automotive systems, we see our approach in this contribution as an additional architecture viewpoint and requirement refinement process.
Our results can later be used in the existing framework to model safety as an orthogonal or emergent property of a system in several architectures.

\subsection{System Architectures \& Performance Monitoring}

Functional system architectures are mostly based on the information processing of human beings. 
Thus, architectures are a representation of functionalities, which a system provides in order to fulfill its purpose.
Rasmussen \cite{rasmussen_skills_1983} divides the human processing regarding the granularity of behavior and access to the necessary knowledge. 
Donges \cite{donges_driver_2016} derives a three layer model of human driving behavior based on Rasmussens` skills, rules, and knowledge layers.

The necessity for performance monitoring in hierarchical systems \cite{mesarovic1979}, AI-systems in general \cite{passino1989} and automated vehicle systems \cite{dickmanns1987}, respectively, has already been formulated several decades ago.
As a consequence, Maurer and Dickmanns \cite{maurer_system_1997} propose a functional system architecture for vision-based autonomous road vehicle guidance, which was explicitly extended by performance monitoring by Maurer \cite{maurer2000}.
In the architecture every module contributing to fulfill the driving task shall provide quality measures which are aggregated in a performance monitoring and assessed during behavior decision.
The practical examples from the demonstrating vehicles VaMoRs and VAmP distinguish meta information and quality measures for performance monitoring.
Meta information is collected as heartbeats and message counters to determine if sensors and software processes are running.
Quality measures are presented for the execution of control algorithms in terms of transient response, overshoot width and control deviation.
Performance monitoring is also classified as a \emph{safety executive} design pattern which indicates that the monitoring is part of a safety concept which was not further defined in the thesis.

Pellkofer \cite{Pellkofer2003} and Siedersberger \cite{Siedersberger2003} concretize the concepts of Maurer:
They further define \emph{capabilities} needed for the driving task and propose a technical framework for monitoring \emph{agents} which execute capabilities (Siedersberger) and use the resulting measures in the behavior decision (Pellkofer).
To model and analyze hierarchical dependencies of capabilities regarding the driving task they introduce the so called \emph{capability net}.
The hierarchy starts from abstract behavior (e.g. free driving) to the implementation narrow capabilities in the system (e.g. control lateral movement or perceiving lanes).

Reschka et al. \cite{reschka2012surveillance} propose a surveillance system based on performance criteria for the (Automation Level 2) research vehicle Leonie in the project \emph{Stadtpilot}.
Performance criteria for the system are identified based on functional degradation strategies which are part of the safety concept for the vehicle.
Since Level 2 automation relies on a human supervisor as a fall-back strategy, the demanded state in case of performance degradation is a request for takeover.

Bergmiller \cite{bergmiller2015towards} uses the concept of capabilities to describe and monitor the performance of the over-actuated by-wire demonstration vehicle \emph{MOBILE}.
The work details movement capabilities with concepts of sports psychology and introduces a fuzzy based framework to determine the actual performance of actuation system and outlines the possibilities of functional redundancies in case of actuator errors or failures.

To summarize previous work, we identified a semantic gap between the abstract concepts of capabilities for fulfilling a driving mission \cite{maurer2000}, the first technical concepts \cite{Pellkofer2003, Siedersberger2003}, as well as the fuzzy-logic-based framework for performance calculation \cite{bergmiller2015towards} for functional safety.
This semantic gap can be addressed by behavioral safety requirements based on a functional hazard analysis in a development process for Level 3 and 4 automation.
Therefore, we presented an approach how safety requirements can be mapped to a knowledge based system to determine the actual performance compared to behavioral safety requirements in the limited use case of an unmanned protective vehicle \cite{nolte_towards_2017}.
In this contribution we will describe how the proposed and implemented concepts of the previously mentioned authors can be embedded in a top-down development of requirements and how these requirements can be allocated in corresponding architectural views.

A singular perspective on functionality is not sufficient.
Especially when it comes to developing an autonomous vehicle, a deeper review of constraints and requirements (e.g. regarding performance) is necessary for structuring an overall system. 
Coste-Mani\`ere and Simmons \cite{coste-maniere_architecture_2000} states that architectures are the foundation of designing, implementing and validating robotic systems. 
Depending on the needs of the different applications provided in a system, different architectures are necessary. 

Regarding safety goals and the resulting requirements, functionalities provided by a system have to be defined. 
Becker et al. \cite{becker_architecture_2015} discuss requirements for maintaining a safe state of a vehicle at any time. 
For that reason, the system needs to detect an occurring fault and, afterwards, transfer the vehicle into a predefined safe state. 
The duration of this process depends on a variety of parameters, e.g. the type of automated function and the driving conditions. 
As a prerequisite for a fall back mode, the vehicle has to be equipped with redundancies in case of degraded functionality. 
This applies to every component implementing a specific function (e.g. the communication network) and depends on the degree of relevance for meeting the safety goals.
However, the relevant components are not visible in a singular architecture.

Jo et al. \cite{jo_development_2015} describe a development process for automated vehicles. 
The process includes a basic structure (following the ``sense-plan-act'' pattern) and holds a ``system management'' module, which, among other things, detects faults occurring in the system and, if necessary, changes the driving mode into ``manual''. 
Hence, the fault management algorithm monitors the health status of the overall system and degrades its functionality if the system reaches an unsafe state. 
In addition, a software architecture and its modules are derived from the basic structure and afterwards the implemented modules are mapped to a distributed component architecture. 
Although the perspectives are linked, there is no holistic architecture framework. 
Hence, requirements and constraints between the architectural perspectives are not visible and therefore not traceable.

Despite the early presented contributions on architectures and performance monitoring, Tas et al. \cite{tas_functional_2016} propose a functional system architecture, which includes concepts for monitoring  system performance and can thus provide information about the current reliability of the functionality w.r.t. the system. 
Besides a ``sense - plan - act'' structure, each functional model contains additional modules for monitoring, which is similar to the demanded quality measures in the presented approach of Maurer \cite{maurer2000}.
Eventually, the information about the performance gathered in the single components is fused within a ``system performance assessment'' block. 
As the performance assessment is located orthogonal and seemingly independent from the focus on information flow in the architecture, this contribution gives a hint that a functional architecture does not cover the concerns of performance monitoring and reliability in a system.  

Bach et al. \cite{bach_taxonomy_2018} propose a taxonomy of upcoming E/E-features and categorize these into \emph{integrated, distributed and cross-linked} features.
The features are then assigned to the categories according to their used functional range in the system.
Integrated features are local functions which are directly coupled to a hardware component like anti-lock braking or engine control.
Distributed features make use of multiple integrated features and combine them to a chain of functions with use of extrospective sensors like lane keeping assist or adaptive cruise control.
The most complex cross-linked features then add a complementary environment model to fulfill parts of the whole driving task like a lane change assist or a highway pilot.
The features are then modeled in a hierarchical logical architecture with a physics, a raw information, a filtered information and an interpreted information level.
The taxonomy covers safety systems, but the resulting architecture only shows information flow, connectivity and logically clustered functions. 

In summary, the introduced approaches highlight different aspects of structuring overall systems. 
In order to develop a safe system, it is indispensable to also define the dependencies (via requirements, constraints etc.) between these perspectives. 

\section{Behavioral safety in architectural views}

In the following, we describe an example scenario and derive a behavioral safety strategy based on an exemplary hazard analysis.
Afterwards, we present how the scenario-specific behavior of an automated vehicle can be expressed in a functional system architecture and a more dynamic capability based view on the system.
Note, that the term behavior in our understanding describes the externally visible system behavior in contrast to most of the contributions in the related work section. 
Previous publications consider the internal system (i.e. software-, or hardware-) behavior while we focus on the externally visible system behavior, more in line with previous own publications \cite{nolte_towards_2017} and Waymo's formulation of \emph{behavioral safety}.

\subsection{Example scenario and safety assessment}

\begin{figure}[htbp]
	\centering
	\includegraphics[width=0.48\textwidth]{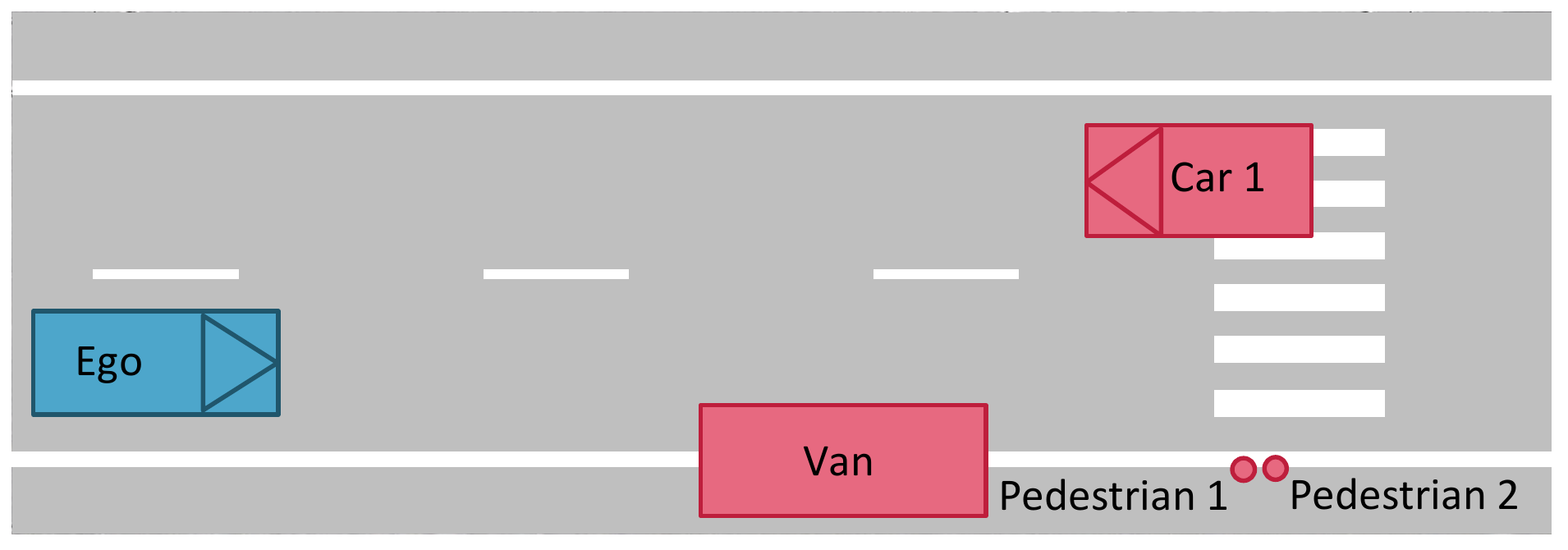}
	\caption{Example scenario at a pedestrian crossing with a parked van and oncoming traffic}
	\label{fig:Scenario}
\end{figure}

Fig.~\ref{fig:Scenario} serves as an example of an inner-city scenario to determine an exemplary safety goal.
We choose a worst-case based risk assessment \cite{khan_use_2001}, based on a language description of a scenario, as this provides a basis for qualitative risk assessment than a quantitative approach starting at a very detailed level.
%The resulting functional safety requirements can not be modeled in a singular architecture as we will show in the following section.
The scenarios evolves as follows:
An automated vehicle approaches a pedestrian crossing at 25 mph.
A van is parking in front of the pedestrian crossing halfway on the side walk.
There is oncoming traffic approaching the automated vehicle and two pedestrians standing at the pedestrian crossing.
The pedestrians are occluded from the point-of-view of the automated vehicle.

While there are multiple possible accidents in this particular scenario, we will focus on the accident:

\noindent\emph{\textbf{A-1}: Deadly injuries to pedestrians at a pedestrian crossing}.
The externally visible behavior of the automated vehicle can be divided into lateral and longitudinal behavior.
With these categories we can systematically derive deviations from the demanded safe behavior to identify hazards.
The hazards regarding longitudinal behavior can be formulated to:

\noindent\emph{\textbf{H-1}: Automated vehicle does not react to pedestrians.}\\
\emph{\textbf{H-2}: Automated vehicle approaches ped. crossing too fast.} \\
\emph{\textbf{H-3}: Automated vehicle does not stop in front of pedestrian crossing with pedestrians present.}

According to Reschka and Maurer, an automated vehicle operates in a safe state when 'the vehicle is driving automatically within its functional boundaries, especially with safe speed and adequate safety distances.' \cite{reschka_conditions_2015}.
To mitigate some of the identified hazards, we can state the externally visible behavior required to maintain a safe state as the safety goal:

\noindent\emph{\textbf{SG-1}: Approach pedestrian crossing with adequate speed.}

The example scenario holds many vaguely described components such as weather, exact speeds and distances, and intentions of other traffic participants.
In real world traffic, many variations of a single scenario can be encountered.
This makes it impossible to derive concrete values as technical requirements at this stage of the process for all variations of a given scenario.
An approach to address the complexity arising from all possible variations, is a step-by-step process, resolving these vague descriptions.

In a first step, we formulated the safety goal also containing the vague term \emph{adequate}, which subsumes several technical requirements.
By determining influences which resolve the term adequate to concrete values, decisions for safe behavior are moved from the design phase to runtime decision making.
In order to enable and argue safe behavior, we need to model influences on decision making as well as interdependencies of these influences in an explicit manner and allocate it to architectural views which represent the concerns of safety requirements.
This concept gives the system the ability to act self-aware within its functional boundaries which are identified and elaborated in the following by looking at a possible worst case in our example scenario: Pedestrians suddenly emerging from the occluded space.

Lin \cite{lin_robot_nodate} motivates thought experiments for situations which are not very likely to happen, but still possible to occur.
Many of these experiments are ``no win scenarios'', which cannot be resolved correctly or in a completely safe manner (such as the trolley problem).
For the given scenario, one of the worst-case expectations would be a pedestrian suddenly appearing in front of the vehicle as soon as it enters the occluded pedestrian crossing, even at low speed.
The most conservative driving strategy in our scenario would thus be to reduce speed while approaching the pedestrian crossing and come to a full stop in front of the crossing for every possible variation of the scenario.
A human driver, in contrast, will at some point (or distance) decide to pass the pedestrian crossing at an individually chosen speed, reflecting the personally accepted risk\footnote{This implies, that each driver will chose a different speed based on her or his personality and current state of mind.}.

Our example scenario contains inner city speed levels, where stopping a vehicle is not a very unusual behavior and does not compulsorily result in an unsafe state for the passengers in the automated vehicle.
For this reason, we formulate the risk minimal state (RMS) for passing the pedestrian crossing below a certain distance, which only leaves mitigation strategies as a choice:

\emph{\textbf{RMS}: Come to complete stop.}

\begin{figure}[htbp]
	\centering
		\includegraphics[width=0.85\columnwidth]{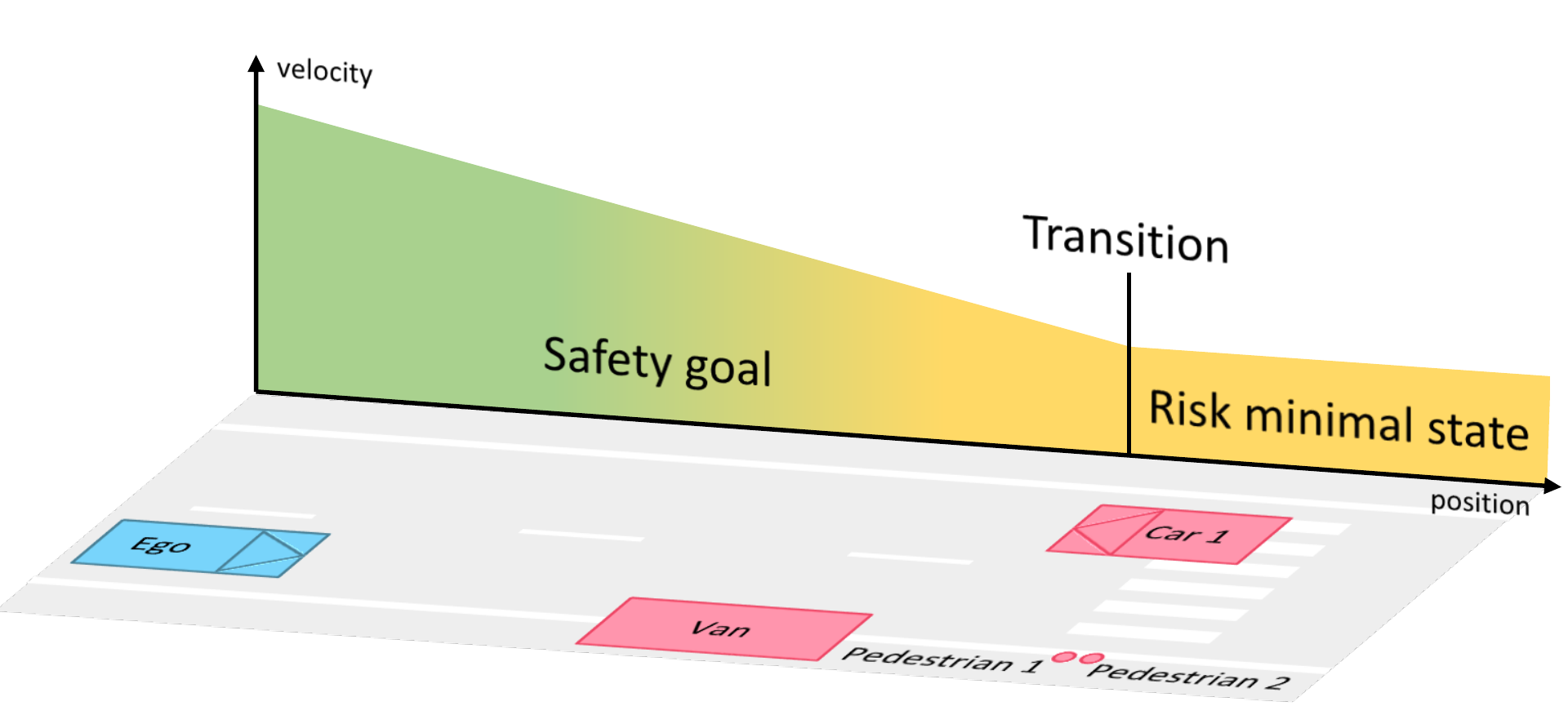}
	\caption{Relationship between safety goal and risk minimal state}
	\label{fig:rms}
\end{figure}

As automated vehicles shall fulfill the human's driving task, we must explicitly determine the boundary (in this case distance and speed) between the safe state by fulfilling the safety goal and a state which only allows mitigation strategies in order to fulfill the driving mission.
The boundary (or point of transition to mitigation-only strategies) has to be determined by an acceptable risk which is not yet commonly defined.
However, when considering a release process for automated vehicles, this boundary must be explicitly modeled in the design process and has to be documented for traceable arguments of safety and remaining risks.
Fig.~\ref{fig:rms} shows the relationship of the safe state by fulfilling the safety goal and the risk minimal state over the distance of the vehicle to the pedestrian crossing.

As introduced, we want to achieve safety in the aspect of behavioral safety but also implement the resulting requirements functionally safe.
Functional safety mainly deals with E/E-failures which can occur at any time during the operation of the vehicle. 
This is the main reason why functional safety can be expressed and treated through failure rates and not as behavioral safety with scenario-dependent driving strategies.
Some of the failures such as power related issues make it impossible to maintain a safe driving state as demanded by the safety goal.
Thus, a transition to the risk minimal state is also the fall-back strategy for E/E-failures where degradation is not applicable.

Since the transition to the risk minimal state can be implemented by the braking system with a fail safe state of \emph{brakes closed}, we will further explain the development of possible strategies to achieve the safety goal.
For this case, the term \emph{adequate} depends on several aspects of the current driving scenario but also on the system state in terms of a system health.
Exemplary aspects are explained in the following, consider also \cite{kelly_rough_1998} for additional metrics:

\subsubsection{Current detection range}

In the best case, the current detection range is similar to the maximum range which can be derived from the data sheets of the environment sensors and possible restrictions of this range due to processing steps.
In the worst case, the detection range can be severely impeded by external influences.
Hasirlioglu and Riener \cite{hasirlioglu_introduction_2017}, for example, recently presented the influence of weather effects on automotive surround sensors.
In case of sensor failures, the whole system will have to degrade in performance as the environment and other traffic participants may only be measured with higher uncertainty.
To describe the performance of the current detection range and quality metrics such as standard deviations of measured states of perceived objects (e.g. dimensions, relative position, velocity, etc.), and the explicit representation of visible areas in the current scene can be used.

\subsubsection{Maximum deceleration and minimum stopping distance}

The possible maximum deceleration and resulting minimum stopping distance on the one hand depends on the maximum applicable brake pressure, as well as the tire state (e.g. temperature), and the mass distributed in the vehicle.
%In case of electric vehicles additional brake force may applied by engine braking.
In addition to these vehicle-related properties, external influences, such as the road condition, determine the friction coefficient between the road surface and the vehicle's tires  impact the minimum stopping distance.
As a result, in order to determine an adequate speed to approach a pedestrian crossing, the vehicle needs to know its own capability to brake but also the road conditions in the current scenario.

\subsubsection{Unknown objects in the current scene}

A major influence on the determination of the term adequate is the treatment of uncertainty in the perceived scene.
Position and motion of objects around the ego vehicle can only be determined with limited accuracy and are always subject to noise.
In addition, parts of the environment may be occluded for optical sensors, while it could be assumed, that objects can emerge from every such occluded area.
This leads to the necessity of making assumptions in terms of worst-case predictions for the  movement of visible objects and existence of occluded objects \cite{nolte_representing_2018}.
The assumptions made in this context determine the vehicle's behavior significantly, depending on how conservative the predictions are.
Hence, the determination of adequate speed in the scenario (c.f. exemplary safety goal) depends on the acceptable risk the system is allowed to take.

To conclude the exemplary aspects which influence the chosen adequate speed in our scenario, the vehicle needs a self perception as part of a safety concept, besides a scene representation containing explicitly formulated occlusions and free space.
Only by combining both aspects, safe decisions and a trade-off between risk and safety can be achieved by determining the boundary for the transition into a risk minimal state.
In the following, we show how these aspects can be modeled in a functional system architecture, which represents the structure and information flow, but also in an capability based view on the system, which builds the model for a self perception.

\subsection{Functional system architecture}

\begin{figure}
	\centering
	\includegraphics[trim=0cm 2cm 0cm 0cm, width=.95\columnwidth]{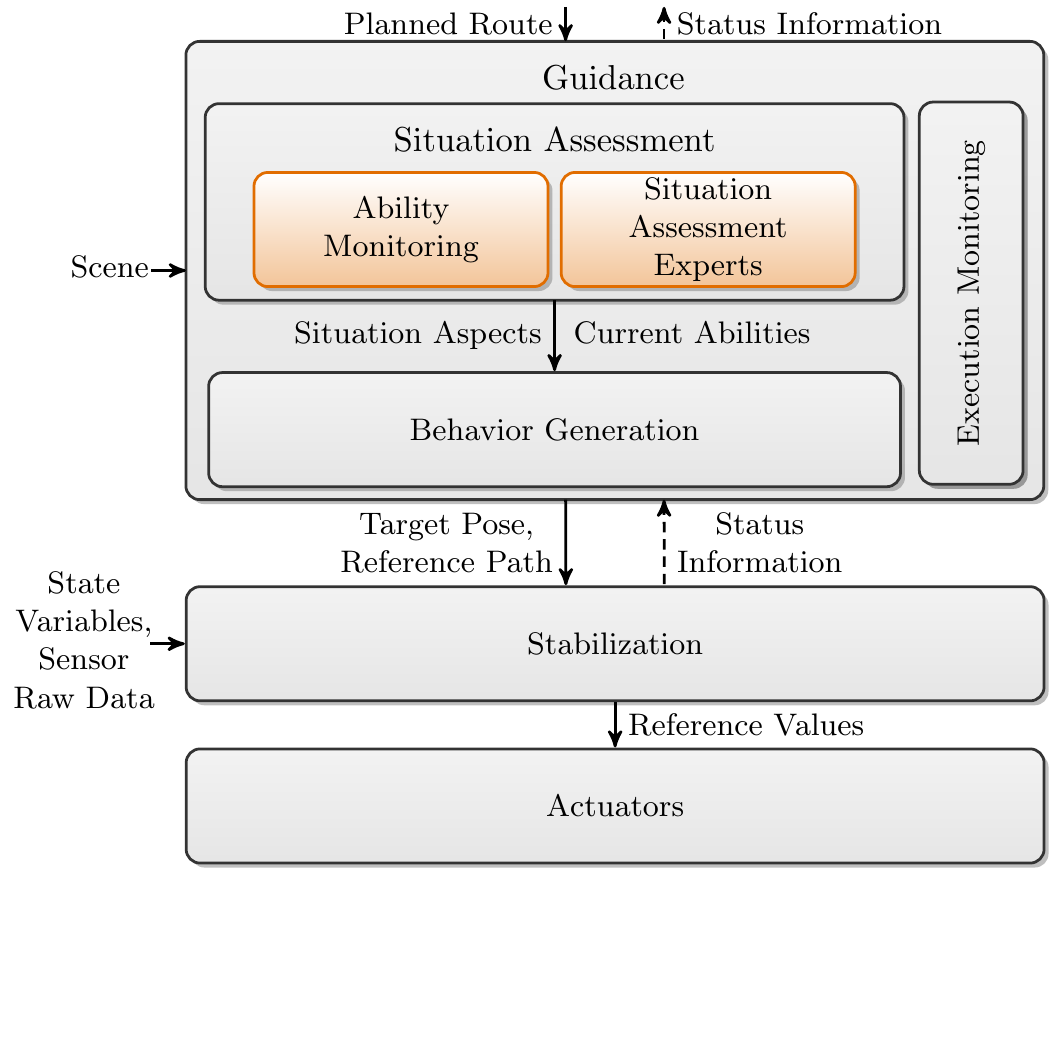}
	\caption{Extract from a functional architecture based on \cite{matthaei_autonomous_2015,nolte_model_2017}, focusing on the tactical level.}
	\label{fig:fusysarch}
\end{figure}

Fig.~\ref{fig:fusysarch} shows a part of our functional system architecture based on the initial work of Maurer \cite{maurer2000} and Matthaei and Maurer \cite{matthaei_autonomous_2015} and a refinement for the use of a model predictive control approach for trajectory planning of Nolte et al. \cite{nolte_model_2017}.
Since functional architectures are a very common view on systems, we briefly want to introduce the ``visible'' aspects of a self-perceiving safety concept.
Input to the guidance and the situation assessment block is the scene which is aggregated from sensor processing, model based filtering and augmentation of objects with semantics from the infrastructure.
The guidance block holds the functional components for behavior decisions on the tactical level.
This is the most central point to influence the externally visible behavior which implements our requirements of behavioral safety.
Based on Nolte et al. \cite{nolte_model_2017}, we separate the functional components in the guidance block to \emph{situation assessment experts\footnote{Experts are meant as functional components which are capable of assessing parts of the whole driving task.}} and \emph{capability monitoring}.
The functional system architecture shows data flow, thus it visualizes which information both components receive and the output generated towards behavior generation.
This structural viewpoint does not change for different scenarios and thus requirements regarding performance cannot be analyzed or modeled further in this view on the system.

\subsection{Self-perception as part of a safety concept}
What changes between scenarios, however, is the required system performance for fulfilling the system's mission:
A tightly packed maneuvering space with moving objects and occluded areas e.g. requires more precise perception, localization and control, compared to a highway scenario.
Accurately assessing the system's performance is a non-trivial task, because of the huge number of interdependencies between functions, as well as software, and hardware components in a system.

On the other hand, influences on a functions's performance cannot directly be derived from data flow in the functional architecture.
Consider a lateral control task as an example: While its performance on a behavioral level depends on the quality of the generated set points, the control performance also depends on the available actuators:
Inaccuracies in the context model, e.g. caused by a lack of performance in the environment perception system, manifest in inaccurate set points for trajectory generation and thus control.
On the other hand, failures in the actuation system cause a degraded control performance.

While the former dependency can be derived from information flow, the latter dependency can not be easily derived from the functional architecture.
For this reason, we have extended the work of \cite{Pellkofer2003} and \cite{Siedersberger2003} by introducing skill \& ability graphs \cite{nolte_towards_2017, reschka_ability_2015} for applications in the development process and at runtime.
These graphs explicitly model external system behavior and the necessary dependencies for performance assessment in terms of dependent capabilities which are required to fulfill the system's mission.
\begin{figure}[htbp]
	\includegraphics[width=0.95\columnwidth]{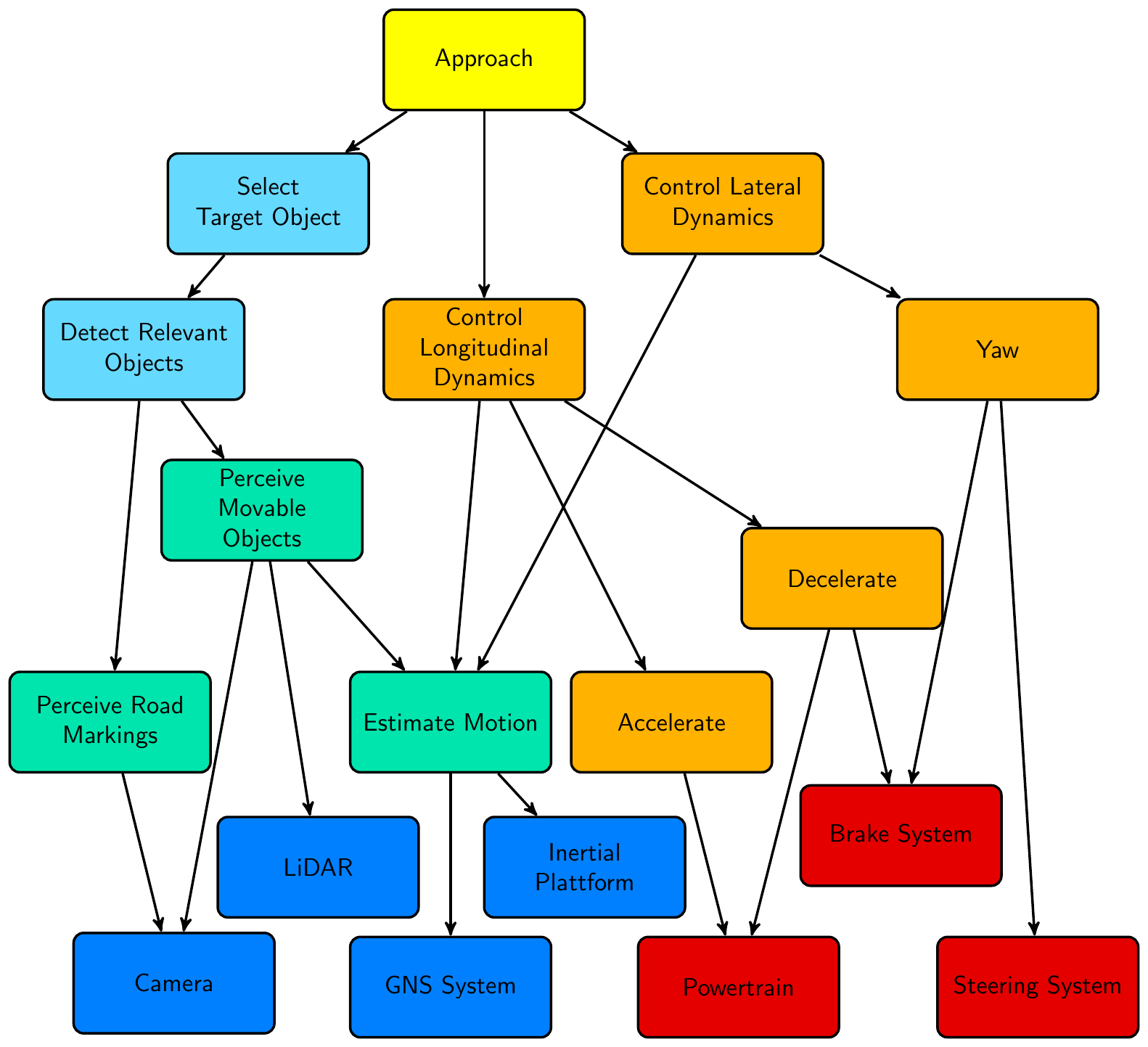}
	\caption{Skill graph for the given scenario: Arrows depict a \emph{requires} relationship.}
	\label{fig:ability}
\end{figure}

For comprehensive performance monitoring, data from large parts of the system is required (cf. \cite{tas_functionalperformance_2017}), while the information generated from the monitors is mainly utilized for decision making.
Considering this and the partial independence of performance indicators from information flow in a functional architecture motivates the introduction of an additional capability viewpoint.
We place the dependency graphs for performance assessment in this viewpoint, complementing the existing architectural framework presented by \cite{broy_toward_2009}.
A skill graph showing the required skills for approaching the crosswalk in the given example scenario is depicted in Fig. \ref{fig:ability}.

As part of a safety concept, the introduction of this additional capability viewpoint provides several benefits:
The skill nodes can be detailed from a behavior to a technical level.
By annotating requirements to the skill nodes, the graph can assist in detailing the formulated coarse functional requirements to technical requirements.

\begin{figure}[tbhp]
	\includegraphics[width=0.98\columnwidth]{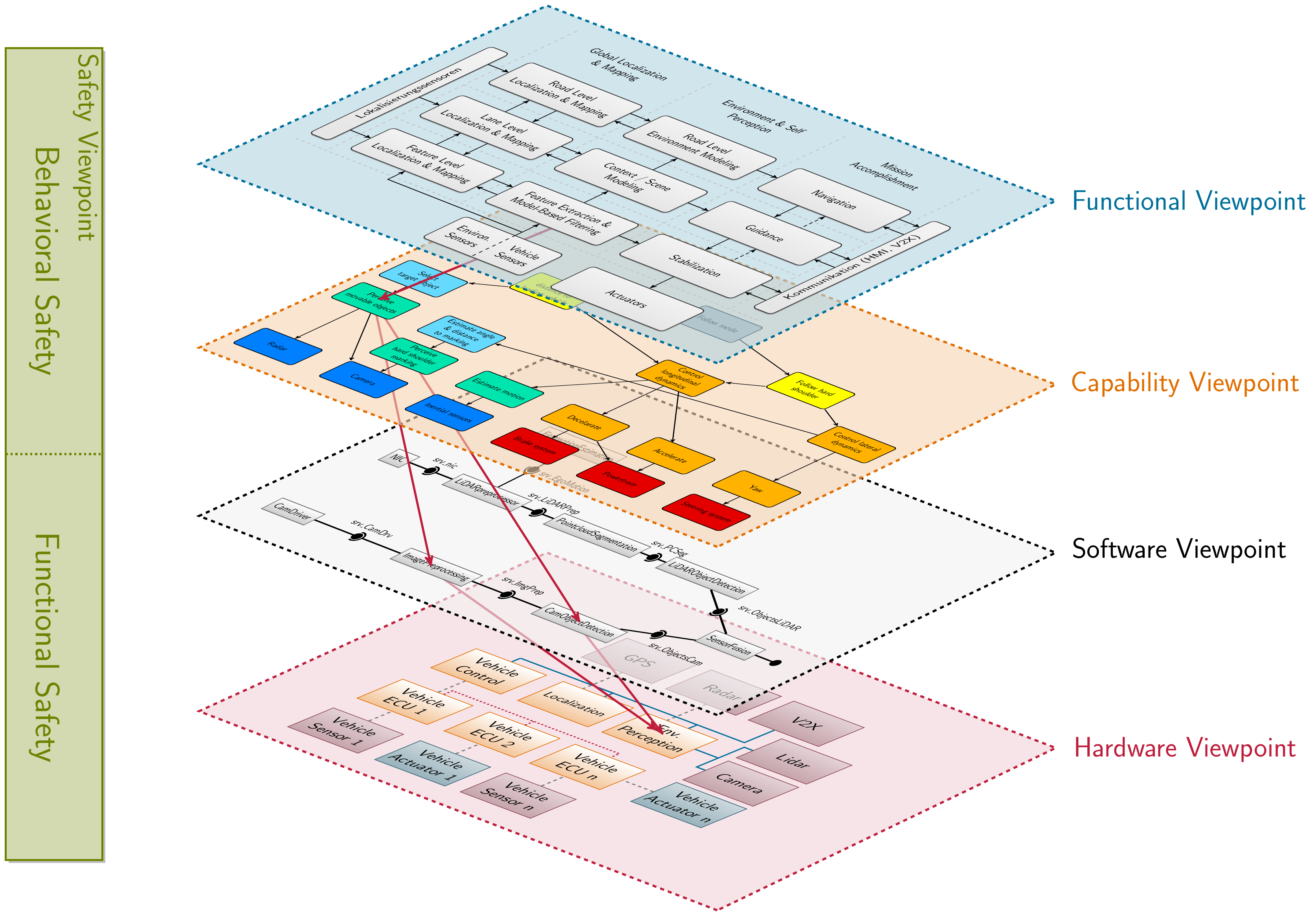}
	\caption{Architecture framework, extended by a capability viewpoint. A safety viewpoint is added as our motivating example for a cross-cutting viewpoint. Concerns of behavioral safety are addressed in functional and capability viewpoints, classic functional safety is reflected in software and hardware viewpoint. Red arrows indicate exemplary mapping relations, i.e. correspondences, between viewpoints.}
	\label{fig:architecture_framework}
\end{figure}

%\todo{example?}
Performance metrics for runtime monitoring can then be derived from annotated requirements.
We consider the capability viewpoint an intermediate representation of (cf.~Fig.~\ref{fig:architecture_framework}) a functional and a software viewpoint (e.g.~represented in a component architecture).
Formulating correspondences and correspondence rules around the capability viewpoint eventually enables tracing the system's external visible behavior to software behavior or hardware properties, as will be described in the following.

Establishing traceable and consistent relations between architectural viewpoints has been widely discussed in the past (cf. \cite{callow_addressing_2011, dajsuren_formalizing_2014}). 
A promising approach in this context is to use correspondence rules as guidelines to formulate model transformations between different viewpoints (\cite{callow_addressing_2011,dajsuren_formalizing_2014,schlatow_towards_2017}), thus providing a formalized way of storing correspondences.
These formal model transformations are often represented as graph transformations (\cite{dajsuren_formalizing_2014,schlatow_towards_2017}), e.g. as $n:m$ mappings between nodes and edges, while nodes represent (functional, software, hardware, etc.) components and edges represent interfaces.
For formal definitions of such transformations, refer to \cite{dajsuren_formalizing_2014,schlatow_towards_2017}.
The component transformations are indicated as red arrows in Fig.~\ref{fig:architecture_framework}, implying that a functional block can be mapped to specific capabilities, which can in turn be mapped to software components, which are assigned to hardware components.
The traceability towards the system's external behavior demanded above can then be reached by additionally annotating the formulated requirements to the respective nodes and edges in the graphs.

\section{Conclusion and Future work}
In this paper, we discussed the necessity of architecture frameworks for automated vehicles from a safety perspective with behavioral safety as a major concern.
The introduction of an additional capability viewpoint in an architectural framework assists the decomposition of functional requirements into technical requirements in the development process.
Performance monitoring presents a possibility to perform a trade-off related to the risk involved in the automated vehicle's actions.
By including system performance in decision making, premature design decisions about acceptable risk can be moved from the design phase to runtime.

Future research will on the one hand focus on further integrating the presented architecture framework into development processes and evaluate traceability from the concept phase of an ISO-related design process to test and validation.
It is also to be determined, how the specification of the different viewpoints holds in iterative design processes such as presented by \cite{graubohm_systematic_2017}.
An additional track of research will aim at formalizing the necessary model transformation around the capability viewpoint.

\section*{Acknowledgment}
This work was partially funded by the DFG under funding number FOR 1800 Conrolling Concurrent Change and the project aFAS funded by the German Federal Ministry of Economics and Technology (BMWi).

\IEEEtriggeratref{25}
\bibliographystyle{IEEEtran}
\bibliography{ITSC2018}

\end{document}